\documentclass[a4paper,12pt]{amsart}
\usepackage{amssymb,amsmath,mathrsfs}
\usepackage[usenames,dvipsnames]{xcolor}
\usepackage{hyperref}
\usepackage{tensor}
\usepackage[left=1.2in,top=1in,right=1.2in,bottom=1in,headheight=0.8in,foot=0.5in]{geometry}
\usepackage[nodisplayskipstretch]{setspace} 
\usepackage{mdframed}
\usepackage{comment}

\hypersetup{
	colorlinks=true,         
	linkcolor=MidnightBlue,          
	citecolor=MidnightBlue,
	urlcolor=MidnightBlue            
 }

\usepackage{natbib}
\setcitestyle{aysep={}} % author date

\title{\sc Spacetime Conventionalism Revisited}
\date{Draft of \today.}

\usepackage{amsaddr}

\author{Ufuk I. Tasdan}
\address{\vspace{-0.8pc}University of Bristol\\Bristol, BS8 1TH, United Kingdom}
\email{\href{mailto:iu.tasdan@bristol.ac.uk}{iu.tasdan@bristol.ac.uk}}

\author{Karim P. Y. Th\'ebault}
\address{\vspace{-0.8pc}University of Bristol\\Bristol, BS8 1TH, United Kingdom}
\email{\href{mailto:karim.thebault@bristol.ac.uk}{karim.thebault@bristol.ac.uk}}

\makeatletter
\let\uppercasenonmath\@gobble

\newcommand{\citeA}{\citet}

\newtheorem*{bach}{Bach Conjecture}

\begin{document}
\setstretch{1.2}
\maketitle

\begin{abstract}
We provide five rearticulations of the thesis that the structure of spacetime is conventional, rather than empirically determined, based upon variation of the structures that are empirically underdetermined and modal contexts in which this underdetermination occurs. Three of the five formulations of conventionalism will be found to fail. Two are found to open up new interesting problems for researchers in the foundations of general relativity. In all five cases, our analysis explores the interplay between geometric identities, symmetry, conformal structure, and the dynamical content of physical theories with the conventionalism dialectic deployed as a tool of explication, clarification, and exploration.      
\end{abstract}
\tableofcontents
\setstretch{1.4}
\section{Conventionalism About What?}
\subsection{Geometric Conventionalism and Universal Forces}
Geometric conventionalism is a form of underdetermination thesis regarding the space or spacetime structure of the physical theory.\footnote{The thesis of conventionalism regarding the geometry of space was first proposed by \citet{poincare1902science}, following in the tradition of \citet{riemann1854ueber}, and developed in a variety of forms by \citet{dingler1911grundlagen}, \citet{carnap1922dr}, and \citet{reichenbach1928philosophy}. See \citet{ben2006conventionalism} and \cite{duerr2022reichenbach}  for an overview. \citet{ivanova2015conventionalism,ivanova2015conventionalisma} gives a helpful analysis of Poincar{\'e}'s conventionalism, and \citet{torretti1978hugo} of Dingler's. A translation of the 1919-1921 doctoral dissertation of \citet{carnap1922dr} can be found in  \citet{carus2019rudolf}. Further analysis with varying degrees of historical grounding can be found in \citep{grunbaum1968geometry,lawrence1974space,sklar1985philosophy,glymour1977epistemology,friedmanfoundations,malament1985modest,norton1994geometry,weatherall2014geometry,dewar:2022}.} 
One may demarcate varieties of geometric conventionalism to arise from how one answers the following two questions:
\begin{enumerate}
    \item Which geometric statements regarding a model of space (or spacetime) are not empirically-determined truths? 
    \item Which physical differences arise in the comparison of different conventions about such statements?
\end{enumerate}
 \citet{reichenbach1928philosophy} famously responded to the first question that statements regarding the spatial or spacetime \textit{metric tensor} of any given model are not empirically determined; and, to the second question that the physical difference between two metrics (and thus two models) would manifest as an apparent \textit{universal force}. In particular, \citet{reichenbach1928philosophy}, like \cite{poincare1902science} in the disc experiment, compares the measuring rod of observers living in two different worlds having flat spacetime. In this thought experiment, inhabitants of one of the world have rulers under the effect of heat and perceive the spacetime as curved. However, ``(h)eat as a force can thus be demonstrated directly" (p. 13) because heat affects different materials differently. As a result, the geometry of spacetime is underdetermined if the heat is taken as a universal force:
\begin{quotation}
``Thus the only distinguishing characteristic of a field of
heat is the fact that it causes different effects on different materials.
But we could very well imagine that the coefficients of heat expansion
of all materials might be equal then no difference would exist between
a field of heat and the geometry of space.'' \citep[p. 26]{reichenbach1928philosophy} 
\end{quotation}

What forms of conventionalism might we consider beyond the specific form articulated in terms of universal forces?  As just noted, in his original argument Reichenbach took metrical relations of spacetime to be conventional and made the specific assumption that the choice in convention could be underdetermined by introducing universal forces. Significantly, \citet{dieks1987gravitation} argues that it is in the Reichenbachian spirit to interpret gravitation as a universal `force', since any effect of the curvature of spacetime can be explained by a suitable gravitational effect. Indeed, in \citet{reichenbach1928philosophy} we find the following text:
 \begin{quotation}
 [The] universal effect of gravitation on all kinds of measuring instruments defines therefore a single geometry. In this respect we may say that gravitation is \emph{geometrized}.
\citep[p. 256]{reichenbach1928philosophy} (italics in original)
\end{quotation}
An explicit defence of this more liberal `universal effect' reading can be found in the later work of \cite{carnap1966p}:
\begin{quotation}
Reichenbach called them `forces', but it is preferable here to speak [...] in a more general way, as two kinds of `effects'. (Forces can be introduced
later to explain the effects.) [p. 169]
\end{quotation}
Further discussion of this point in the context of the interpretation of Reichenbach can be found in \cite{sexl:1970} and \cite[\S2.4.2]{duerr2022reichenbach}.

In what follows we will draw inspiration from these ideas for re-framing of the debate via generalised forms of the conventionalist thesis that do not rely on force-based representations. In particular, we will consider different formulations of the spacetime conventionalist thesis in terms of \textit{physical differences} that are taken to result from different proposals for the relevant \textit{universal effect}.

\subsection{Rearticulating Spacetime Conventionalism}

We have just seen that one is able to historically situate a mildly generalised form of geometric conventionalism based upon universal effects rather than universal forces. For a foundational, as opposed to primarily historical, perspective the next step is then to consider the salience of wider forms of conventionalist thesis.\footnote{Our project is not primarily a historical one and so we will not seek to defend the historical as opposed to foundational salience of our generalisations. That said, our framework for generalisation overlaps in several core respects with the reconstruction of Poincar\'e's conventionalism provided by \cite[p.167]{duerr2022reichenbach}.} To do this productively, however, one must place some constraints upon the relevant argument pattern. With this in mind, we will consider the following generalised forms of our two questions:
\begin{enumerate}
    \item Which statements regarding the structure of a spacetime theory are empirically underdetermined? 
    \item Which physical differences arise in the comparison of different conventions about such structure?
\end{enumerate}
Setting the two questions up in this way allows us to confine ourselves within a generalised spacetime conventionalist argument pattern whilst allowing for rearticulation along various dimensions. 

The first dimension of rearticulation we will pursue involves considering various different ways of formulating the relevant physical differences in (2). The second dimension of rearticulation is to vary the structure in question in (2). Such structure should, of course, still be physically well-motivated as the basic structure of a spacetime theory, but need not necessarily be a geometric structure. Third, and finally, we can identify a more subtle dimension of rearticulation in terms of the  \textit{modal context} of empirical underdetermination thesis. The idea is that the context in which one is answering the questions (1) and (2) is itself underdetermined to the degree that it might be specified as obtaining between: i) two observers within a single model of a theory; ii) two models within the same theory; iii) two models of different theories; or iv) two ways of interpreting the sets of models within a single theory. What is essential within our family of generalised notions of spacetime conventionalism is that in each and every case it is required that a basic structure of a spacetime theory is empirically underdetermined, and this underdetermination leads to the possibility for physical differences to arise between conventions regarding how to break the underdetermination.

In what follows we will explore a selection of sights of interest in the conceptual space defined along these three dimensions of rearticulation. Some of the pathways towards spacetime conventionalism that we will explore will take us rather far from the original debates, and others will ultimately be found to lead us into a \textit{cu-de-sac}.  We take this tour to be a valuable and productive one for two principal reasons. 

First, in what follows,  we will deploy our generalised framework for varieties of spacetime conventionalism as a tool of foundational analysis towards both clarification of old conceptual problems, or indeed pseudo-problems, and the identification of new problems. Articulation of new conceptual problems can, in turn, be understood as a constitutive element of scientific progress according to the framework articulated by \cite{laudan1977progress}, c.f. \cite{feyerabend:1981}. In this sense, our ultimate aim is to apply conventionalism as a lens to advance the scientific understanding of spacetime theory. 

Second, we take our project to provide important clarificatory insights into the formal interplay between geometric identities, symmetry, conformal structure, and dynamical equations within the foundations of general relativity. The five forms of conventionalism articulated in what follows are woven together by twin formal threads of conformal structure and the Bianchi identities. Each of the forms of conventionalism we will consider, to at least some degree, trades on a package of re-reading the interconnection between parts of the formalism of general relativity based on one or both of these aspects. We will consider putative conventionalisms relating to non-conformal structure, local conformal structure and to the decomposition of the Einstein equation into conformally invariant and non-conformally invariant parts. Further, we will consider putative conventionalisms based upon understanding the role of the Bianchi identities as formal geometric identities, conservation equations, and dynamical field equations.       

We start in Section \ref{Sec1} by considering a formulation of geometric conventionalism due to \cite{weatherall2014geometry}. We consider the implications of the no-go result proved by the authors in the context of a general framework for the analysis of the methodological implications of no-go theorems due to \cite{dardashti:2021}. We will find that this analysis of the Weatherall-Manchak no-go result further motivates the programme for rearticulation along a different dimension that we have just set out.

In Section \ref{Sec2} we present our first attempt at rearticulation of conventionalism in which we focus on inertial structure and putative universal effects that result from differing representations of local stress-energy 4-current by coincident observers following different world-lines. Ultimately, the view is found to be unconvincing, since there are good reasons to reject the physical salience of the energy 4-current and thus this prima facie plausible form of conventionalism can be dissolved.  In Section \ref{Sec3} we then consider the conformal structure and the universal effect of tidal deformation. We argue that by considering the invariant decomposition of the Riemann tensor into Weyl and Ricci parts we can dissolve a further potential source of conventionalism with regard to the deforming and non-deforming aspects of tidal effects. In each case, although the new forms of conventionalism are found to fail, our discussion serves to elucidate new perspectives on the representational roles of different objects within the foundations of general relativity. 

In Section \ref{Sec4}, we pull our focus back to a wider form of interpretational question regarding the dynamical understanding of the Weyl tensor fields and the role of the Bianchi identities. In particular, we consider a putative underdetermination with regard to the \textit{nomic structure} of general relativity. This analysis serves to open up the possibility for a very different form of conventionalism in terms of how to isolate the dynamical content of general relativity. In particular, we argue that underdetermination with regard to the structure that specifies which relativistic spacetimes are dynamically possible models leads to conventionalism with regard to which fields are understood as dynamical and which non-dynamical. Furthermore, we find that consideration of the initial value problem of general relativity serves to further establish, rather than dissolve, the putative underdetermination with regard to nomic structure that is at the heart of Spacetime Conventionalism 4. 

Finally, in Section \ref{Sec5}, we consider a proposal based upon the decomposition of the Einstein tensor into a conformally invariant and non-conformally invariant part. The potential for underdetermination of this decomposition is then taken to be the hallmark of a novel form of conventionalism whose validity is expressible in terms of (the failure of) a precise mathematical `Bach conjecture', whose truth is as yet unknown.

\section{Non-Conformal Structure and Universal Forces}
\label{Sec1}
\bigskip
\begin{mdframed}
\noindent
\textbf{Spacetime Conventionalism 1}
\begin{enumerate}
    \item The \textit{non-conformal structure} of \textit{relativistic spacetime} is not an empirically-determined truth. 
    \item Physical differences \textit{between conformally equivalent models of relativistic spacetime} with regard to the existence of \textit{universal forces} arise in the comparison of different conventions about \textit{affine structure}.
\end{enumerate}
\end{mdframed}
\bigskip

Any pair of conformal equivalent relativistic spacetimes are, by definition, such that $(M,g_{\mu\nu})$ and $(M,\tilde{g}_{\mu\nu})$ where $\tilde{g}_{\mu\nu}=\Omega^2 g_{\mu\nu}$. Each of these spacetimes is accompanied by a metric compatible, torsion-free, Levi-Civita derivative operator, $\nabla$ and $\tilde{\nabla}$ respectively. Relativistic spacetime conventionalism can then be formulated as the thesis that the \textit{same curve}, $\gamma$, can be a geodesic relative to $\nabla$ but be associated with an acceleration relative to $\tilde{\nabla}$ given by a `universal force' induced by a tensor field $G_{\mu\nu}$ via the equation $G^\nu_{\:\mu} \tilde{\xi}^\mu$ where $\tilde{\xi}$ is the tangent field to $\gamma$ with unit length relative to $\tilde{g}_{\mu\nu}$. So formulated a relativistic conventionalist thesis, equivalent to Spacetime Conventionalism 1 has a contradiction as a deductive consequence  \cite[Proposition 2]{weatherall2014geometry}. 

The result of Weatherall and Manchak constitutes a `no-go' theorem for a specific form of spacetime conventionalism and so understood, the theorem might be taken to close off one of the principal avenues for the articulation of such a view within the context of spacetime theory. It is worth noting in this context that recent analysis due to \cite{duerr2022reichenbach} has strongly contested whether the characterisation of conventionalism within the antecedent conditions of the Weatherall and Manchak no-go result is historically or conceptually adequate. Whilst our perspective on spacetime conventionalism intersects in a number of respects with Duer and Ben-Menahem, our attitude to the work of Weatherall and Manchak is rather different. This difference can be understood by drawing upon the hugely insightful work on no-go theorems in physics due to \cite{dardashti:2021}.

Within Dardashti's framework, a no-go result has been established if and only if an inconsistency arises between a derived consequence of a set of physical assumptions, $P$, represented by a mathematical structure $M$, and a goal $G$  within a framework, $F$. By contraposition, the implications of the no-go result can then be understood in terms of different `methodological pathways' based upon a disjunct of the form $\neg P \vee \neg M \vee \neg F \vee \neg G$. Presuming that we do not want to give up the goal, we may then reinterpret the  implication as a `go theorem' with the implication $\neg P \vee \neg M \vee \neg F$. That is, if we would like to achieve a certain goal, then the theorem provides us with a formal basis to articulate the specific mathematical, physical, or  framework assumptions one must reject to achieve it.

Returning to the Weatherall-Manchak result. Since the theorem contained in their Proposition 2 is both valid and non-trivial, we take there to be good cause to explore its implications as a `go theorem' in the context of the negation of the various physical, mathematical and framework assumptions.\footnote{See \cite{duerr2022reichenbach} for specific arguments that involve questioning the Weatherall-Manchak result on the basis of both its mathematical assumptions, (NORM) that $\tilde{\xi}$ is a unit norm and (RIEM) that $(M,\tilde{g}_{\mu\nu})$ is a Riemannian geometry, and physical assumptions, (CONF) that the old and new metrics are conformally related $\tilde{g}_{\mu\nu}=\Omega^2 g_{\mu\nu}$ and (FORCE) that the `universal force' is induced by a tensor field $G_{\mu\nu}$ via the equation $G^\nu_{\:\mu} \tilde{\xi}^\mu$.} That is, we take the result to motivate `methodological pathways' in which we seek to articulate different forms of the generalised spacetime conventionalist goal based upon different physical, mathematical and framework assumptions. Our specific approach is based upon articulating a set of such pathways in terms of the different dimensions of rearticulation of conventionalism as per the analysis of the introduction. Thus, our project is further motivated by a `go theorem' interpretation of the Weatherall-Manchak result. We start by considering a form of spacetime conventionalism founded upon inertial structure and an intra-model modal context.

\section{Inertial Structure and Local Stress-Energy Flux}
\label{Sec2}
\bigskip
\begin{mdframed}
\noindent\textbf{Spacetime Conventionalism 2}
\begin{enumerate}
    \item The \textit{inertial structure} of a \textit{particular relativistic spacetime} is not an empirically-determined truth. 
    \item Physical differences between \textit{coincident observers} with regard to \textit{local stress-energy flux} arise in the comparison of different conventions about which \textit{time-like vector field} is chosen as an observer's inertial frame.
\end{enumerate}
\end{mdframed}
\bigskip
The second form of the conventionality of spacetime conventionality we will consider relates to the interpretation of \textit{different} observers of the \textit{same model}, and thus the \textit{same geometric facts}, in terms of \textit{different local dynamical facts}. The idea is that within a spacetime there can be observers that are coincident and yet will disagree about the dynamical interpretation of the same local geometric facts. The formal property that we will isolate in the privileged class of observers who have the analogue of a vanishing universal force is that they are \textit{Killing observers} and thus the answer to the first of our two conventionalism questions is \textit{inertial structure}. The physical differences that feature in the second question are then with regard to local stress-energy flux. 

We can better understand the motivation for Conventionalism 2 by briefly considering the problem of the local definition of conservation of energy in general relativity.\footnote{See \cite{pitts:2010,pitts:2016,pitts:2021,lam2011gravitational,read2020functional,dewar2018gravitational,durr2019ain,duerr2019against} for discussions of the problem of defining gravitational energy. A further noteworthy paper, along lightly different lines, is \cite{lehmkuhl2011}. There it is argued that in the context of general relativity energy-momentum density as expressed by $T_{\mu\nu}$, cannot, in fact, be regarded as an intrinsic property of matter, but rather should be understood as a relational property that matter possesses only in virtue of its relation to spacetime structure. For an outstanding general discussion of energy conditions in general relativity see \cite{curiel2014primer}.} The problem can be straightforwardly stated as follows. For a vanishing cosmological constant, the Einstein equations take the familiar form:
 \begin{equation}
 \label{einstein}
 G_{\mu\nu} = R_{\mu\nu} -\frac{1}{2} R g_{\mu\nu} = 8\pi T _{\mu\nu}
 \end{equation}
where $T _{\mu\nu}$ is the energy-momentum tensor associated with all matter fields and their interactions, and $G_{\mu\nu}$ is the Einstein tensor which is determined by the metric $g_{\mu\nu}$ via the Ricci curvature $R_{\mu\nu}$ and Ricci scalar $R$. 

The Levi-Civita derivative,  $\nabla_\mu$,  can be explicitly constructed from the metric and an arbitrarily chosen derivative operator. Given such a derivative operator on a manifold $M$, it can be proved that there exists a unique smooth tensor field on $M$ such that for all smooth fields $\xi^\sigma $:
\begin{equation}
R^\rho _{\sigma \mu\nu} \xi^\sigma  = -2 \nabla_{[\mu} \nabla_{\nu]} \xi^\rho
\end{equation}
\citep[Lemma 1.8.1]{malament2012topics}. $R^\rho _{\sigma \mu\nu}$ is then the unique Riemann curvature tensor field associated with  $\nabla_\mu$. The Ricci curvature tensor is defined via the contraction of the Riemann curvature as:
\begin{equation}
R_{\sigma \mu}= R^\rho _{\sigma \mu\rho}
\end{equation}
An important identity that holds for any Riemann tensor is the second Bianchi identity. For the torsion-free case, this can be expressed in terms of the Levi-Civita derivative as:
\begin{equation}
\label{Bianchiuncon}
 \nabla_{[\alpha} R^\rho _{\mid \sigma\mid \mu\nu]} =0   
\end{equation}
This equation follows from the basic mathematical properties of the Riemann tensor and is thus naturally understood in this context as a formal identity without physical content.\footnote{See \cite[p.69-79]{malament2012topics} and \cite[p.61]{Landsman:2021} for proofs.} Interestingly, however, when combined with the definition of the Einstein tensor the Equation (\ref{Bianchiuncon}) can be shown to imply:\footnote{See \cite[p.322]{malament2012topics}. N.b. one also uses further symmetry properties of the Riemann tensor in this short derivation.}
\begin{equation}
    \nabla_\mu G^{\mu\nu} = \nabla_\mu T^{\mu\nu} =0
\end{equation}
At this point, it is very tempting to read the equation $\nabla_\mu T^{\mu\nu} =0$ as an expression of the local conservation of energy (or stress-energy) and it seems like we have derived a dynamical conservation equation from a formal identity. However, such a na\"ive way of thinking about energy conservation in GR immediately runs into trouble. 

Consider an observer moving along the integral curve of an arbitrary time-like vector field $\xi^\mu$. Let us define the \textit{stress-energy 4-current} along the direction defined by this vector as $j[\xi^\mu]=T_{\:\nu}^{\mu} \xi^\nu$. This 4-current \textit{does not} in general vanish, so we cannot be guaranteed that $\nabla_\mu j[\xi^\mu] =0$ by the fact that $\nabla_\mu T^{\mu\nu} =0$. This might be interpreted to mean that we no longer have `strict' conservation of energy according to the curved spacetime version of Gauss' law \cite[pp. 69-70]{wald1984general}. Moreover, we might plausibly interpret $j[\xi^\mu]$ as the conserved current associated with the observers' time translations and thus understand $\nabla_\mu j[\xi^\mu] =0$ to imply a failure of conservation of locally measured energy. Next, recall that the class of Killing vector fields are given by the satisfaction of the Killing equation:
\begin{equation}\label{Killing}
\nabla_\mu \xi_\nu + \nabla_\nu \xi_\mu =0
\end{equation}
This is equivalent to the statement that the metric is invariant along the integral curves of $\xi_\nu$ or, in terms of the Lie derivative, that we have that $\mathfrak{L}_\xi g_{\mu\nu}=0$. The existence of a time-like Killing vector field is then (in this sense) equivalent to the symmetry of time translation invariance holding along the relevant time-like curves. The existence of a time-like Killing vector field is then necessary and sufficient for the vanishing of the stress-energy 4-current. This, in turn, implies that for the class of \textit{Killing observers} whose, worldlines are integral curves of the vector fields solving (\ref{Killing}), the vanishing of the stress-energy 4-current and thus `strict' conservation of energy \textit{is} guaranteed. 
 
 We can then formulate Spacetime Conventionalism 2 by interpreting the stress-energy flux for the non-Killing observers as playing the functional role of a Carnapian universal effect. Killing and non-Killing observers can be coincident within the same spacetime patch but will attribute different interpretations to the same geometric facts in terms of the presence or absence of this additional associated flux. Putative physical differences with regard to local stress-energy flux thus can be understood to arise in the comparison of different conventions about which timelike vector field is chosen as an observer's rest frame. The existence or not of such effects depends upon a convention as to the choice of timelike vector field as a rest frame. 
  
 Three clear lines of response to this argument for Spacetime Conventionalism 2 are available. The first is simply to note that in the general class of Einstein spacetimes Killing observers are not guaranteed to exist at all. The existence or not of Killing vector fields depends upon there being non-trivial continuous isometries of the metric, and there are good reasons to believe that such transformations do not exist in general for physically realistic spacetimes.\footnote{Following the results of \citeA{Fischer:1970}, what can be proved explicitly is that for globally hyperbolic Einstein spacetimes which admit an initial value formulation, the topology of the space of physically distinguishable initial states (i.e. the `superspace' composed of distinct Riemannian three-geometries) is such that the geometries admitting continuous isometries are singular points. Furthermore, initial data without non-trivial Killing vector fields is generic in the sense that `geometries of high symmetry are completely contained in the boundary of geometries of lower symmetry' (p. 303). There is thus a good formal basis to expect that a generic (globally hyperbolic) Einstein spacetime will lack Killing vector fields. At a simpler, more intuitive, level it is not hard to convince oneself that any physically realistic spacetime will be too `messy' in terms of inhomogeneities and anisotropies to contain Killing vector fields.} In a generic spacetime we can therefore expect that there will not be any set of `freely falling' observers who measure locally vanishing stress-energy 4-current. Spacetime Conventionalism 2 would then be best understood as a viable but highly limited conventionalist position.\footnote{It is true that, in general, local conservation of stress-energy will approximately hold for `small enough' patches of spacetime, with how small depending on the relevant curvature \cite[pp. 70]{wald1984general}. Thus, there will, in general, exist `approximate Killing observers' for whom within some `small enough' region there are no tidal forces and gravitational stress-energy 4-current vanishes.  Operationally, however, \textit{observers cannot be arbitrarily small}, and thus making such a move would violate the spirit of the Reichenbachian argument.} 

The second response relies upon the introduction of further structure into the theory. The first step is to note that  the \textit{magnitude} of the stress-energy 4-current depends upon the affine properties of the world-line of the observer. This means that the \textit{quantity} given the flux into a region is still conventional in the sense that it will be different for different observers moving within the same spacetime patch irrespective of the existence of not of Killing vector fields. In this context, one could consider introduction of a gravitational-energy-momentum tensor $t^{\mu\nu}[g_{\alpha\beta},\nabla_{\alpha\beta}]$ which is such that a total energy-momentum complex is conserved with respect to the flat metric's torsion free derivative operator  \citep{pitts:2010,pitts:2016,pitts:2021}. This would eliminate any potential for conventionalism with regard to total energy conservation but at the cost of introducing extra background or auxiliary structures into the theory.   

This brings us to our third ultimately more satisfactory response to Killing conventionalism: the rejection of the physical salience of the energy 4-current. The key observation is that it is far from clear that focusing on localised energy fluxes is germane to the context of relativistic physics; arguably the ambiguities we have encountered are a product of focusing on concepts and quantities that are a relic of the bygone, pre-relativistic era. Consider in particular the fact that \textit{the failure of the vanishing of the stress-energy 4-current can occur for non-Killing observers even in flat spacetimes}. This surely indicates that we should not think of it as encoding the results of  additional universal effect resulting in the failure of energy conservation. Rather it motivates us to look for a completely general characterisation of the \textit{causal basis} of genuine universal gravitational effects due to curvature without the need for background structure: this is to re-frame the discussion to focus on tidal forces and their explicit geometric representation, rather than energy conservation. To do this it will prove essential to consider the physical significance of conformal structure within general relativity. 

\section{Conformal Structure and Tidal Deformation}
\label{Sec3}

\bigskip
\begin{mdframed}
\noindent
\textbf{Spacetime Conventionalism 3}
\begin{enumerate}
    \item The \textit{local conformal structure} of a \textit{particular relativistic spacetime} is not an empirically-determined truth. 
    \item Physical differences between \textit{coincident observers} regard to \textit{mechanical stress due to tidal forces} arise in the comparison of different conventions about the decomposition of the Riemann tensor into deforming and non-deforming parts.
\end{enumerate}
\end{mdframed}
\bigskip

Consider a smooth one-parameter family of geodesics. Define two vector fields: a timelike vector field $\xi^\mu$ tangent to the family of geodesics and a second vector field $\chi^\nu$ that represents the infinitesimal displacement to an infinitesimally nearby geodesic. For a given Riemann curvature tensor, the acceleration due to \textit{geodesic deviation} is given by:
\begin{equation}\label{geodesicdeviation}
a^\mu = -R^\mu_{\beta\nu \alpha}\xi^\alpha\xi^\beta \chi^\nu
\end{equation}
In general terms, tidal `forces' should be understood as the effects of geodesic deviation as induced by Riemann curvature. There are two physically distinct senses in which a geodesic can undergo deviation. The first is through conformal re-scaling, which would change local spatial scales but preserve local shape. The second is through deformation which would preserve local spatial scales but change local shape. 

Each effect can have nontrivial physical consequences. However, there are compelling physical reasons to expect the manifestations of the effects for local observers to be rather different. Since geodesic deviation through conformal re-scaling does not change the shape of a material body it will not induce mechanical stress upon that body. Rather, physical manifestations will be indirect and contingent on the material content and relevant scales and couplings constants. For example, redshifts or thermal effects due to expansion. By contrast, geodesic deviation through deformation will change local shape and thus manifest as a `universal effect' which induces mechanical stress on a material body. We have direct empirical evidence that such stresses are able to overwhelm internal forces in realistic physical situations through the observed phenomena of \textit{spaghettification} in which the deforming geodesic deviation induces an extreme form of tidal shear in the region outside a black hole event horizon that rips stars into thin strips. 

Spacetime Conventionalism 3 is then the view that seeks to exploit underdetermination in the representation of tidal forces. The proposal would be that different coincident observers may decompose the consequences of the geodesic deviation given by Equation \eqref{geodesicdeviation} differently, and so adopt different conventions with regard to the degree of tidal deformation and associated mechanical stress. This would not only underdetermine the tidal `forces' in a physically significant sense but also underdetermine the local conformal structure of spacetime. 

The proposed new formulation of spacetime conventionalism has, however, been set up to fail. It is precisely in the relevant respect that the Riemann curvature tensor has a unique decomposition in terms of its Ricci and Weyl parts. This decomposition equates to exactly the decomposition of geodesic deviation into deforming and non-deforming parts. We can see this as follows. The Weyl curvature tensor, $C_{\rho\sigma\mu\nu}$ can be defined via the Riemann and Ricci curvature tensors as the traceless tensor:  

\begin{equation}
C_{\rho\sigma\mu\nu}= R_{\rho\sigma\mu\nu}-\frac{1}{2}\left(g_{\rho[\nu}R_{\mu]\sigma}+g_{\sigma[\mu}R_{\nu]\rho}\right)-\frac{R}{6}\left(g_{\rho[\mu}g_{\mu]\sigma}\right)\label{eq:rieman expansion1}
\end{equation}    
The tensorial nature of this expression of course means that the decomposition of Riemann curvature into Ricci and Weyl parts is invariant under the push-forward of diffeomorphisms, $\phi$ of the spacetime manifold $M$. That is, for any three tensors which are related such that $A=B+C$ we will have that $\phi^\star(A)=\phi^\star (B+C) =\phi^\star B+\phi^\star C$. Observers using different coordinate systems in a given spacetime will necessarily agree on the decomposition. Moreover, the decomposition of the Riemann curvature tensor into Ricci and Weyl components has its origin in the symmetry transformation properties of the Riemann tensor. In particular, we can explicitly derive the decomposition by considering the symmetric and anti-symmetric parts of a decomposition of the Riemann tensor in terms of SO(3,1) irreducible tensors.\footnote{Following \citeA{ramond1997field} we can decompose $R^\rho _{\sigma \mu\nu}$ in terms of $SU(2)\otimes SU(2)$ as it is locally isomorphic to $SO(3,1)$. We decompose the Riemann curvature tensor into symmetric parts

\begin{equation}
\left(3\otimes3\right)\oplus\left(5\otimes1\right)\oplus\left(1\otimes5\right)\oplus\left(1\otimes1\right)\oplus\left(1\otimes1\right)\label{eq:decomp}
\end{equation}

and antisymmetric parts

\[
\left(3\otimes3\right)\oplus\left(3\otimes1\right)\oplus\left(1\otimes3\right)
\]

We then find that the Weyl tensor $C_{\rho\sigma\mu\nu}$ transforms as $\left(5\otimes1\right)\oplus\left(1\otimes5\right)$
and it is conformally symmetric. The object that transforms as $\left(3\otimes3\right)$ in the
symmetric part the corresponds to traceless part of Ricci tensor $R$.} We thus find that there is a solid mathematical foundation for taking the distinction as observer-independent and intrinsic to the structure of a given spacetime. 

A further decomposition of the Weyl tensor then allows an explicit and insightful connection to tidal deformation.\footnote{Here we are following  \citeA{goswami:2021}. The origin of this decomposition is \cite{matte_1953}. Further discussion can be found in \cite{bel2000}. Thanks to Juliusz Doboszewski for help with these historical sources.} Let us first interpret the Weyl tensor as the free gravitational field, and the metric tensor as its (2nd order) potential field and consider a timelike unit vector field $\xi^\nu \xi_\nu=-1$ representing a family of  observers. Next, parallel to the way one can split the electromagnetic field into electric and magnetic parts in the rest frame of $\xi^\nu$, we can split the Weyl curvature tensor, $C_{\rho\sigma\mu\nu}$, into electric and magnetic parts constructed as symmetric traceless tensors orthogonal to $\xi^\nu$. The explicit expressions are:
\begin{eqnarray}
\label{weyldecomp}
E_{\rho\sigma}=C_{\rho\sigma\mu\nu}\xi^\mu \xi^\nu \\
H_{\rho\sigma}=\frac{1}{2} \epsilon_{\rho\nu\alpha}C^{\nu\alpha}_{\sigma\mu}\xi^\mu
\end{eqnarray}
where $\epsilon_{\rho\sigma\mu}$ is the effective volume element in the rest space of the co-moving observer and is explicitly given by:
\begin{equation}
    \epsilon_{\rho\sigma\mu} = \sqrt{|\text{det g}|} \delta^0_{[\rho} \delta^1_{\sigma} \delta^2_{\mu}\delta^3_{\nu]}\xi^\nu
\end{equation}\citep[Appendix A]{goswami:2021}. The equation for geodesic deviation due to the Weyl curvature is then simply:
\begin{eqnarray}
a^\mu &=& -C^\mu_{\beta\nu \alpha}\xi^\alpha\xi^\beta \chi^\nu  \\
&=&   E^\mu_{\nu} \chi^\nu
\end{eqnarray}
We can thus understand the electric part of the Weyl tensor as uniquely responsible for the mechanical stress associated with the tidal deformation effect -- that is, the geodesic deviation that changes the shape of bodies in geodesic motion.\footnote{There is also an important connection between the  electric part of the Weyl tensor and gravitational waves - see \citeA{goswami:2021} for more details on the formal relation. A recent, insightful discussion of gravitational waves and gravitational energy in the context of the analogy with electromagnetic waves can be found in \cite{Gomes:2023}.} Tidal deformation is a non-conventional effect which can be associated with a specific part of the Weyl curvature tensor of any given spacetime. Spacetime Conventionalism 3 is a resounding failure. 

\section{Nomic Structure and Dynamical Fields}
\label{Sec4}
\bigskip
\begin{mdframed}
\noindent
\textbf{Spacetime Conventionalism 4}
\begin{enumerate}
    \item The \textit{nomic structure} that specifies which \textit{relativistic spacetimes} are dynamically possible models is not an empirically-determined truth. 
    \item Physical differences with regard to \textit{dynamical and non-dynamical} interpretations of field equations arise in the comparison of different conventions about whether certain \textit{fields are understood to couple or interact} within equations.
\end{enumerate}
\end{mdframed}
\bigskip
Whilst the dynamical role of the Weyl tensor forecloses the possibility for a putative conventionalism about tidal deformation, it opens up an opportunity for a more interesting, and ultimately more tenable, form of conventionalism with regard to interpretations of the equations that enforce the connection between Weyl and Ricci curvature. In this section, we will consider a novel form of conventionalism that is built upon underdetermination with regard to the nomic or law-like structure that specifies which relativistic spacetimes are dynamically possible models. In particular, an underdetermination with regard to which equations are taken to encode nomic structure that directly restricts dynamically possible models and which equations are taken to be gauge identities or constraints that restrict kinematically possible models. This underdetermination then, in turn, leads to conventionalism with regard to which fields are understood to couple or interact which in turn implies \textit{physical differences}. We can articulate the line of thought behind this form of conventionalism as follows.

First, recall that solving the Einstein equation (\ref{einstein}) \textit{does not} uniquely determine the Riemann curvature. This can be seen most obviously in the example of vacuum Einstein spacetimes which are defined as the class of Ricci flat Lorentzian manifolds -- i.e. Einstein spacetimes in which the Ricci curvature tensor $R_{\mu\nu}$ is zero. The Riemann curvature of a Ricci flat Lorentzian manifold is entirely determined by the Weyl curvature tensor $C_{\mu\nu\alpha\beta}$. In contrast, we might also consider Weyl flat spacetimes, in which the Ricci curvature encodes all geometric degrees of freedom. Weyl flat spacetime cannot admit any local deformations of shape and are illustrated most vividly by the FLRW spacetimes that approximately describe our universe. The intersection of the Ricci flat and Weyl flat cases is then the unique Riemann flat spacetime: Minkowski spacetime. In general, a spacetime will have non-zero Ricci and Weyl curvature and the obvious question is then how these two forms of curvature can be related if the Einstein equation only relates to the Ricci part.   

The answer is found in the Bianchi identities. Equation  (\ref{Bianchiuncon}) can be re-written as:
\begin{equation}
\label{Bianchi2}
\nabla_{[\alpha} R_{\mu\nu]\rho\sigma}=0  
\end{equation}
this can be expanded via the Weyl-Ricci decomposition (\ref{eq:rieman expansion1}) as:
\begin{equation}
\label{Bianchi}
\nabla_\nu C^{\rho\sigma\mu\nu}= \nabla^{[\sigma}R^{\rho]\mu}+\frac{1}{2}g^{\mu[\sigma}\nabla^{\rho]} R
\end{equation}
In the context of the further decomposition of the Weyl tensor as per Equation (\ref{weyldecomp}), the contracted Bianchi identities then lead directly to the temporal and spatial derivatives of the electric and magnetic part of the Weyl tensor, and these equations then lead in vacuo to the propagation equations for gravitational waves \citep{goswami:2021}.\footnote{For the original treatments along related lines see \cite{newman:1962a, newman:1962b, hawking:1966}.} 

The next crucial step towards Spacetime Conventionalism 4 is to consider the form of Maxwell's equations:
\begin{equation}
\nabla_\sigma F^{\rho\sigma} = J^{\rho}   
\end{equation}
where $F^{\rho\sigma}$ is the electromagnetic field tensor and $J^{\rho} $ is the source current, and note that we can re-write the equations (\ref{Bianchi}) such that they take an analogous form: 
\begin{equation}
\label{source1}
\nabla_\nu  C^{\rho\sigma\mu\nu} = J^{\rho\sigma\mu}  
\end{equation}
where $J^{\rho\sigma\mu}= \nabla^{[\sigma} R^{\rho]\mu}+\frac{1}{2}g^{\mu[\sigma}\nabla^{\rho]} R$  \cite[p.85]{hawking:1973}. This striking analogy suggests we might think of the Equation (\ref{Bianchi}) as \textit{field equations for the Weyl curvature}, just as the Einstein equations provide field equations for the Ricci curvature. Furthermore, we can use the Einstein equation to re-write the source current directly in terms of stress-energy tensor such that we have an expression of the form:
\begin{equation}
\label{source2}
\nabla_\nu  C^{\rho\sigma\mu\nu} = \nabla^{[\sigma}T^{\rho]\mu}+\frac{1}{2}g^{\mu[\sigma}\nabla^{\rho]} T_\nu^\nu 
\end{equation}
We would then seem to have a dynamical equation for the Weyl curvature as sourced by the stress-energy tensor \citep{danehkar:2009}. In such circumstances, the difference between Weyl and Ricci curvature appears to have disappeared in the sense that both can be seen as \textit{dynamical fields} encoding geometric degrees of freedom that are sourced by stress-energy tensor.  

Notwithstanding the argument just given, we can also arrive at a non-dynamical interpretation of the Weyl tensor by starting with a different reading of the \textit{modal status} of the Bianchi identities. That is, we understand the identities as encoding restrictions on \textit{kinematical} rather than \textit{dynamical} possibilities. A Bianchi identity is, in the general case, defined as the consequence of the `gauge' symmetry properties of a theory via Noether's second theorem.\footnote{For a detailed discussion of Noether's second theorem in a historical context see \cite{kosmann:2010}. For a rigorous formal overview see \cite{Olver:1993}. For discussion in the context of gauge theories and quantization see \cite{Henneaux:1992a}.} The Noether derivation of Bianchi identities does not require stationarity of the relevant action. Then, in the context of general relativity, as was discussed in Section \ref{Sec2}, the second Bianchi identity follows from the basic mathematical properties of the Riemann tensor and is thus naturally understood as a \textit{formal identity without physical content}. As such, Equation (\ref{Bianchi}) is derivable completely independently of the action and would hold for any spacetime theory formulated on Riemannian geometries with a torsion-free connection. 

These considerations lead us to a second non-dynamical interpretational option in which we treat the Bianchi identities as encoding kinematic rather than nomic structure. In the context of general relativity, this means that the space of kinematically possible models of the theory is preselected such that the identities are obeyed. This, in turn, means that the Ricci and Weyl curvatures are kinematically constrained to be coordinated such that the Riemann curvature obeys the relevant identity. Dynamics is then encoded solely within the Einstein equation which then fixes the dynamically allowed Ricci curvature. Residual freedom within the Weyl curvature can be understood as fixed via the choice of initial conditions. Under such an interpretation the relation between the Ricci and Weyl curvature is a product of a pre-established kinematical harmony, not a substantive dynamical relationship. The Weyl tensor is a non-dynamical field. 

A useful way of understanding the connection between the interpretation of equations and the interpretation of fields has been suggested by \cite{lehmkuhl2011}: 

\begin{quote}
[A]n interaction demands that all fields present are dynamical fields [...] it seems sensible to make a distinction between speaking of two fields \textit{interacting} and two fields \textit{coupling}. For a dynamical field can couple to a non-dynamical field [...] but we would not speak of an \textit{interaction} if only one of the two fields was dynamical: a non-dynamical field acts without being acted upon if it couples to a dynamical field. Hence, two fields interacting should be seen as sufficient but not necessary for the fields to couple, whereas two fields coupling is necessary but not sufficient
for the two fields to interact.
(p.469)
\end{quote}
Following this formulation, we would then have it that if the Bianchi identities (\ref{Bianchi}) describe an interaction then the Weyl tensor is a dynamical field of the same status as the Ricci tensor; they each act whilst simultaneously being acted upon. However, if the Bianchi identities (\ref{Bianchi}) describe a coupling, then we should offer a non-dynamical interpretation of the Weyl curvature tensor, it merely couples to the Ricci tensor, the two fields do not interact; the Ricci tensor acts upon the Weyl tensor without being acted upon.\footnote{This idea parallels in certain key respects the discussion of the distinction between matter and spacetime found in \cite{MARTENS2020237}.}

What would appear to be the best strategy for breaking this interpretational underdetermination is to consider the initial value problem.\footnote{Thanks to Henrique Gomes for suggesting this idea to us.} In particular, if the initial value problem of general relativity were found to be fully specifiable independently of the Bianchi identities, one could argue that only the Einstein equations encode nomic structure, and the non-dynamical interpretation is supported. Conversely, if the Bianchi identities play an explicit role in encoding interactions between degrees of freedom within the initial value problem then they would encode nomic structure and the dynamical interpretation would be supported.\footnote{The idea that (well-posed) initial value problems could play a privileged role in picking out the ontic structure of physical theories is explored in a rather different context by \cite{callender:2017}. See also \cite{baron:2021,james:2022}.}     

Approaches to the explicit solution of the initial value problem in general relativity proceed via hyperbolic reductions based upon a particular gauge choice. The most famous of these is the original `harmonic' or `wave' gauge treatment due to Choqouet-Bruhat.\footnote{Here we are following \cite[\S VI]{Choquet:2008} and \cite[7.5]{Landsman:2021}.}  In this approach, although the Bianchi identities do play a substantive role, they do not form part of the basic system of hyperbolic dynamical equations that are obtained as a reduction of the Einstein equations. This would seem to support the non-dynamical interpretation. However, at least in vacuum, there exists an alternative approach to the hyperbolic reduction of general relativity due to \cite{Friedrich:1996} in which the Bianchi identities \textit{are} explicitly understood as hyperbolic propagation equations for the Weyl tensor. Explicitly, and considering the vacuum case, within this approach, the Weyl tensor is treated as \textit{one of the fundamental dynamical variables} in a system of equations given by:
\begin{eqnarray} \label{Weyl_Bianchi}
    R^\mu_{\nu\lambda\rho}  &=& C^\mu_{\nu\lambda\rho} \\
    \nabla_\mu  C^\mu_{\nu\lambda\rho} &=&0 
\end{eqnarray}
which are equivalent to the vacuum forms of the Einstein equation and Bianchi identity respectively. This is precisely to understand the Einstein equations and Bianchi identities as interaction equations of equal status and, moreover, to explicitly treat the Weyl tensor as a dynamical field.\footnote{For a more general application of the idea of treating the Bianchi identities as alternative representations of the field equations in the context of matter systems see \cite{friedrich:2000}.}  

Study of the initial value problem of general relativity thus serves to further establish, rather than dissolve, the putative underdetermination with regard to nomic structure that is at the heart of Spacetime Conventionalism 4. If we appeal to the initial value problem to identify the nomic structure of general relativity then we find that such a specification depends upon a choice as to which approach to hyperbolic reduction we apply. At least in the vacuum case, this then explicitly allows for different options with regard to the attribution of a dynamical or non-dynamical role to the Weyl tensor. 

We can draw an analogy between this situation and the role of the cosmological constant.\footnote{See \cite{MARTENS2020237} for a related discussion regarding modified gravity and dark matter.} Consider a `$\Lambda$-vacuum' spacetime of the form:
\begin{equation}
G_{\mu\nu} + g_{\mu\nu}\Lambda =0
\end{equation}
Writing the equation in this way makes it natural to think of $\Lambda$ as a non-dynamical constant of nature.  We might, of course, alternatively re-write the same equation as:
\begin{equation}
G_{\mu\nu} =- g_{\mu\nu}\Lambda 
\end{equation}
In this context, it is natural to think of $\Lambda$ as a dynamical source for the Ricci degrees of freedom. These are completely trivial rewritings of the same equation, so it is not clear what could hang on the dynamical vs. non-dynamical interpretation of Lambda. However, there are good reasons \textit{coming from outside the theory} to treat the cosmological constant vs. dark energy interpretational question as a substantive physical issue \citep{peebles2003cosmological,huterer2011accelerating}.

This requirement for reference to external physical or physics principles could also be taken to be required for the case of our two interpretations of the Ricci-Weyl relationship. Within general relativity, the equations themselves do not give priority to the differing dynamical and non-dynamical interpretations. However, in a wider theoretical context, and considering, in particular, the relevance to quantization, the distinction between the interpretation may in fact ground genuine physical differences. For example, whereas it is generally the case that kinematical restrictions are converted to super-selection rules in quantization, dynamical restrictions, such as conservation laws, are applied as quantum nomological restrictions, which allows for their violation subject to the uncertainty principle.\footnote{See \cite{gryb:2016} for more discussion in the context of the problem of time and the cosmological constant.} Spacetime Conventionalism 4 might thus offer insight into alternative heuristic strategies for theory extension.

Finally, an alternative strategy for breaking the underdetermination is to consider the deeper mathematical structure behind the tensor fields in question. Appeal to the primacy of such structure as the basis upon which to designate which fields are dynamical and which non-dynamical.  In particular, one can understand the Weyl and Ricci tensors as composite objects built out of the metric tensor, spinors or connection coefficients, and then argue that since the Bianchi identities but not the Einstein equations, follow from the mathematical definition of these objects, only the latter can be understood as nomic structure.\footnote{Thanks to Henrique Gomes for suggesting this argument to us.} In the context of such a view, Spacetime Conventionalism 4 would, in its failure, again provide a valuable heuristic, in this case towards a particular, non-trivial, mathematical ontology of relativity theory. Such an avenue of investigation warrants further consideration.     

\section{The Bach Conjecture}
\label{Sec5}

\bigskip
\begin{mdframed}
\noindent
\textbf{Spacetime Conventionalism 5}
\begin{enumerate}
    \item The \textit{geometric structure} of spacetime is not an empirically-determined truth. 
    \item Physical differences with regard to which tensor fields play the role of geometric structure, arise through different conventions in the decomposition of the Einstein tensor into conformally invariant and non-conformally invariant parts.
\end{enumerate}
\end{mdframed}
\bigskip

The final form of conventionalism we will consider is, in a sense, a marriage of the formulation of geometric conventionalism in the spirit of \citeA{weatherall2014geometry} with the idea discussed above of understanding conformally invariant tensors as representing the geometric structure of spacetime. The view is, however, to our knowledge entirely novel. Furthermore, it leads to a precise mathematical conjecture the truth of which is as yet unproven. 

The basic starting point for this particular strategy is to look to re-express the Einstein equation in a conformally invariant form. That is, seek to show that Einstein's equation is true if and only if,
\begin{equation}\label{eq:bach}
  B_{\mu\nu} = 8\pi T_{\mu\nu} + A_{\mu\nu},
\end{equation}
where $T_{\mu\nu}$ is the matter-energy tensor appearing in Einstein's equation, $B_{\mu\nu}$ is a conformally invariant tensor, and $A_{\mu\nu}$ is a further tensor that is not conformally invariant. So far this is purely a formal exercise in analysis. The interpretation move is then to take $B_{\mu\nu}$ to characterise facts about geometric spacetime structure, which is, by assumption, understood to be conformally invariant. By contrast, the non-conformally invariant objects $T_{\mu\nu}$ and $A_{\mu\nu}$ are taken to characterise a matter-energy field and a further `universal effect', restrictively.\footnote{The interpretation of $T_{\mu\nu}$ as matter is equivalent to matter criterion G of \cite{MARTENS2020237}.}  Clearly, by the theorem of \citeA{weatherall2014geometry}, $A_{\mu\nu}$ will not be expressible as a Newtonian force.

The re-formulation \eqref{eq:bach} we are looking for is indeed possible, by defining $B_{\mu\nu}$ to be the \emph{Bach tensor}\footnote{Named for Rudolf Bach, not the  Baroque German composer. Although, the former was in fact a pseudonym for Rudolf F\"{o}rster. See Appendix \ref{appendix} for further formal details.} which can be expressed without introducing further structure in terms of the Weyl and Ricci tensors as:
\begin{equation}\label{eq:bach_eqn_of_motion}
  B_{\mu\nu} := \nabla^\sigma\nabla^\rho C_{\rho \mu\nu \sigma} + \tfrac{1}{2}C_{\rho\mu\nu \sigma}R^{\rho \sigma}.
\end{equation}
The Bach tensor is conformally invariant in $D=4$ \citep{bach:1921}. Thus, defining $A_{\mu\nu}:=B_{\mu\nu}-G_{\mu\nu}$, we have that for $D=4$ the Equation \eqref{eq:bach} holds if and only if the Einstein equation does, as desired.\footnote{For $D=2$, all relativistic spacetimes are trivial vacuum solutions to Einstein's equation since it can be proved both that the Einstein tensor is identically zero and that the natural candidate for a two-dimensional Weyl tensor is identically zero \citep{fletcher:2018}. For $D=3$, the Riemann curvature tensor is completely determined by the local matter distribution \citep{giddings:1984}, so although there are interesting conformally invariant tensors in three dimensions, such as the Weyl tensor and Cotton tensor, the dynamics of three-dimensional gravity is such that these are constrained to be zero. Thus, Conventionalism for 5 for $D<4$ is uninteresting.}

The idea of a `pure geometry' interpretation of the Bach tensor is noted in the mathematical physics literature. In particular, \cite{bergman2004conformal} offers the observation that: 
\begin{quote}
The Bach tensor is a tensor built up from pure geometry, and thereby captures necessary features of a space being conformally Einstein in an intrinsic way. \cite[p. 18]{bergman2004conformal}
\end{quote}

The pivotal issue in the context of Spacetime Conventionalism 5, is then whether the Bach tensor $B_{\mu\nu}$ is the \emph{unique} conformally invariant two-place tensor, at least up to a multiplicative constant. Adopting the interpretation of geometric structure and matter-energy above: uniqueness would imply that this decomposition allows one to distinguish uniquely between geometric structure ($B_{\mu\nu}$) and matter-energy ($T_{\mu\nu}$) and the `universal effect' ($A_{\mu\nu}$); and, \emph{non}-uniqueness would amount to a sense in which the distinction between geometric structure and matter-energy is conventional.

Fascinatingly, although this question is both mathematically precise and foundationally important, its status appears, as yet, unsettled. The issue at hand is one which can be more generally associated with finding the conformal invariants in the theory of Lorentzian manifolds: 
\begin{quote}
  ``Classically known conformally invariant tensors include the Weyl conformal curvature tensor, which plays the role of the Riemann curvature tensor, its three-dimensional analogue the Cotton tensor, and the Bach tensor in dimension four. Further examples are not so easy to come by.'' \cite[p.1]{FeffermanGraham2012a}
\end{quote}
Thus, although the review of the literature does not reveal any other known conformal invariants in four dimensions, there is no proof of the absence of such alternatives either.

 One might thus propose that the debate over conventionality and non-conventionality, as described above, is not just a matter of philosophical debate, but of settling a precise mathematical conjecture, whose truth is not yet known:
\begin{bach}
  The Bach tensor is the unique (up to a multiplicative constant) conformally invariant rank 2 tensor field on a four-dimensional Lorentzian manifold.
\end{bach}  
If the Bach Conjecture is false, then Spacetime Conventionalism 5 is true; if the Bach Conjecture is true, then Spacetime Conventionalism 5 is false. We thus take the proof or refutation of the Bach conjecture to be an important problem in the foundations of general relativity that warrants considered attention. 

%It is finally worth remarking that the Bach tensor has many interesting properties: for example, it is known to vanish for metrics that are locally conformal to the Einstein metric, and for self-dual metrics \cite{derdzinski1983b}. It would thus seem that on this interpretation of spacetime structure, the Einstein universe is one with no `pure' spacetime structure, in spite of its evidently positive curvature! Moreover, the vanishing of the Bach tensor has also been shown to be equivalent to $(M,g_{ab})$ satisfying the Yang-Mills Equations $D\star R = 0$ for the Cartan normal conformal connection associated with $g_{ab}$, where $R$ is the curvature associated with that connection \cite{knn2003c}. Analogues of it have been explored in higher dimensions as well \cite{GrahamHirachi2005c,FeffermanGraham2012a}. Clearly, there is a wide variety of interesting physical content to explore in seeking to understand this interpretation.

\section{Summary}

One might reasonably judge proposals for the conventionality of spacetime structure in terms of their  respective novelty, interestingness, and plausibility. Our discussion has characterised and critically evaluated five formulations of the conventionality thesis and we hope the reader will agree that all five have proved worthy of consideration. The first three options considered have been argued to be implausible on grounds that vary from direct refutation (Spacetime Conventionalism 3), to provable falsity under certain assumptions (Spacetime Conventionalism 1), to rejection on the grounds of the lack of physical salience of the central undetermined object (Spacetime Conventionalism 2). Nevertheless, each of the three views illustrates important morals regarding the interplay between empirical and mathematical structures within the foundations of spacetime theory. The final two forms of conventionalism we considered to be novel, interesting, and plausible. Moreover, in terms of analysis of the dynamical role of the Bianchi identities (Spacetime Conventionalism 4) and proof or refutation of the Bach conjecture (Spacetime Conventionalism 5), they each open up a new direction of enquiry that we may hope to be profitably pursued in future researches.    

\section*{Acknowledgements}

The form, scope, and content of this paper were transformed through hugely productive interactions with Bryan Roberts. In particular, both the framing of varieties of conventionalism via the two questions and the formulation of the Bach conjecture is due directly to Bryan Roberts. We further greatly profited from detailed written comments and lengthy discussions with Bryan without which the present paper would be a shadow of what it is. The paper has also profited from a number of other detailed discussions that KT has had over the last years with (among others) Erik Curiel, Juliusz Doboszewsk, Sam Fletcher, Henrique Gomes, Sean Gryb, James Ladyman, Dennis Lehmkuhl, Oliver Pooley, James Read, Simon Saunders, Dave Sloan, Jim Weatherall. We are also grateful to online audiences at the DPG conference Heidelberg and the Philosophy of Physics and Chemistry group in Buenos Aires, Argentina, and to Patrick D\"uerr and two anonymous referees for helpful comments. The authors take full responsibility for any errors, omissions or inaccuracies. 

\bibliographystyle{dcu}
\bibliography{conventionalism}

\appendix
\ref{appendix}

\section{Conformal Tensors and the Bianchi Identities}
\label{appendix}
The Bianchi identities can also be related to the Cotton Tensor $\mathcal{C}_{\rho\sigma\nu}$  \citep{cotton1899varietes}
and Bach Tensor 
$B_{ab}$ as follows. Following \citeA{garcia2004cotton}, we can decompose curvature into irreducible representations with respect to the pseudo-orthogonal group for certain dimensions, $n$, as follows:

\begin{eqnarray}
n=1 & \rightarrow R_{\alpha\beta}=0\\
n=2 & \rightarrow R_{\alpha\beta}=Scalar_{\alpha\beta}\\
n=3 & \rightarrow R_{\alpha\beta}=Scalar_{\alpha\beta}+\widetilde{R}_{\alpha\beta}\\
n\geq4 & \rightarrow R_{\alpha\beta}=Scalar_{\alpha\beta}+\widetilde{R}_{\alpha\beta}+C_{\alpha\beta}
\end{eqnarray}

where $\widetilde{R}$ denotes traceless Ricci part of the decomposition
and every tensor is in its Weyl 2-form. For $n=3$, the Cotton tensor
paper which is given by $DR_{\alpha\beta}=DC_{\alpha\beta}+\frac{2}{n-2}\upsilon_{[\alpha}\land\mathcal{C}_{\beta]}=0$
where $D$ denotes exterior covariant derivative and $\upsilon$ denotes
coframe of a Riemannian space of $n$ dimensions.

\begin{equation}
\mathcal{C}_{\rho\sigma\nu}=2\left(\nabla_{[\alpha}R_{\beta]\gamma}-\frac{1}{2\left(n-1\right)}\nabla_{[\alpha}Rg_{\beta]\gamma}\right)
\end{equation}

and for $n=4$, the Bach tensor can be defined as

\begin{equation}
B_{\alpha\beta}:=\nabla^{\mu}\mathcal{C}_{\alpha\mu\beta}+L^{\mu\nu}C_{\alpha\mu\beta\nu}
\end{equation}

where $L^{\mu\nu}$ is called Schouten tensor and given by

\begin{equation}
 L_{\mu\nu}=R_{\mu\nu}-\frac{1}{2\left(n-1\right)}Rg_{\mu\nu}.   
\end{equation}

Although the Bach and Weyl tensors appear to be independent in tensorial form, they can be directly associated with one another via a variational principle. In particular, we can drive the Bach tensor from a Lagrangian composed of Weyl tensors that takes the form:\footnote{see \cite{fiedler1980exact} for the exact solutions.} 

\begin{equation}
L=C_{\rho \mu\nu \sigma}C^{\rho \mu\nu \sigma}+kg^{\mu\nu}T_{\mu\nu}
\end{equation}

where $T_{\mu\nu}$ is the relevant matter field with a non-zero constant. The relevant Euler-Lagrange equations are then Bach's equation of motion:

\begin{equation}\label{eq:bach_eqn_of_motion_matter}
B_{\mu\nu}=kT_{\mu\nu}
\end{equation}

where $T_{\mu\nu}$ is traceless due to conformal invariance of $B_{\mu\nu}$. 

Additionally, considering the vacuum case, the Bach tensor exhibits parallel behaviour to the Weyl tensor in relation to the Bianchi identities given in \eqref{Weyl_Bianchi}\footnote{see \cite[p. 39]{bergman2004conformal} for the discussion.}:

\begin{eqnarray} \label{Bach_Bianchi}
    \nabla^d C_{abcd} &=& K^d C_{abcd} \\
    B_{\mu\nu} &=& 0
\end{eqnarray}

\end{document}